\documentstyle[preprint,aps]{revtex}

\begin{document}

\draft

\title{Non-universal behavior of finite quantum Hall systems as a
result of weak macroscopic inhomogeneities}

\author{I. M. Ruzin\cite{igoraddress},
N. R. Cooper\cite{nigeladdress}, and B. I. Halperin}
\address{Department of Physics, Harvard University, Cambridge, MA
02138, USA.}

\date{September 14, 1995}

\maketitle

\begin{abstract}
We show that, at low temperatures, macroscopic inhomogeneities of the
electron density in the interior of a finite sample cause a reduction
in the measured conductivity peak heights $\sigma_{xx}^{\rm max}$
compared to the universal values previously predicted for infinite
homogeneous samples.  This effect is expected to occur for the
conductivity peaks measured in standard experimental geometries such
as the Hall bar and the Corbino disc. At the lowest temperatures, the
decrease in $\sigma_{xx}^{\rm max}(T)$ is found to saturate at values
proportional to the difference between the adjacent plateaus in
$\sigma_{xy}$, with a prefactor which depends on the particular
realization of disorder in the sample.  We argue that this provides a
possible explanation of the ``non-universal scaling'' of
$\sigma_{xx}^{\rm max}$ observed in a number of experiments. We also
predict an enhancement of the ``non-local'' resistance due to the
macroscopic inhomogeneities. We argue that, in the Hall bar with a
sharp edge, the enhanced ``non-local'' resistance and the size
corrections to the ``local'' resistance $R_{xx}$ are directly
related. Using this relation, we suggest a method by which the
finite-size corrections may be eliminated from $R_{xx}$ and $R_{xy}$
in this case.
\end{abstract}
\pacs{PACS number: 73.40.Hm}

\section{Introduction}

The fascinating property of the quantum Hall effect (QHE) that
initially attracted such great attention to the phenomenon is the
precise quantization of the Hall conductivity $\sigma_{xy}$ at certain
values of the magnetic field.  Most theoretical research has focused
on the properties of the electron system inside these quantized
plateaus in $\sigma_{xy}$. The plateaus have been associated with the
incompressibility of the 2D electron gas, arising either from Landau
quantization, at integer filling factors, or from electron-electron
interactions, at fractional filling factors. The transition regions,
where $\sigma_{xy}$ crosses over between quantized values and the
longitudinal conductivity $\sigma_{xx}$ experiences maxima, have
received less attention. The main factor which inhibits progress in
this direction is the lack of reproducible experimental results on the
inter-plateau regions, despite the impressive stock of data on the QHE
which has been accumulated over the last decade. In addition to the
fact that the general behavior of the QHE depends on the electron
density, temperature and disorder, samples cut from the same substrate
and measured at the same temperature often reveal different
dependences $\sigma_{xy}(B)$ and $\sigma_{xx}(B)$.  This annoying data
dispersion is particularly apparent at low temperatures.\cite{Tsui}
Certain success in obtaining reproducible data has been achieved only
for the critical behavior of the width of inter-plateau regions at low
temperatures.\cite{widthExp} However, as far as the heights or shapes
of the peaks in $\sigma_{xx}$ are concerned, the general impression is
that too many factors are involved to allow any systematic
conclusions.

On the other hand, there do exist a number of theoretical works which
argue that certain universal behavior of the conductivity tensor must
exist at low temperatures in the regions between well-pronounced pairs
of plateaus (critical regime). Kucera and Streda\cite{KuceraStreda}
considered semi-classical single-electron transport in a partially
filled Landau level for a simple model of a periodic long-range
potential. They found that the maximum value of $\sigma_{xx}$ reached
at half-integer filling factors does not depend on either the
magnitude of the potential or the Landau level number $N$ and is equal
to $e^2/2h$. This result was later mapped onto the fractional regime
by using two related approximations of the correlated electron state:
the dirty boson\cite{Kivelson} and the composite fermion
approach.\cite{ChklovskiLee} A different sort of argument for both
integer and fractional regime was presented in
Refs.~\onlinecite{DykhneRuzin} and \onlinecite{RuzinFeng}. Based on
the rather general assumption that, at low temperatures, the electron
system in a critical transition region can be represented by a random
mixture of two quantum liquids with different quantized local Hall
conductivities $\sigma_2$ and $\sigma_1$, it was shown that
$\sigma_{xx}$ and $\sigma_{xy}$ are connected by a universal relation.
The peak height $\sigma_{xx}^{\rm max}$ was found to be equal to one
half of the difference between the Hall conductivities of the adjacent
plateaus $|\sigma_2-\sigma_1|/2$. For the integer peaks, this result
yields the value $e^2/2h$ obtained in Ref.\onlinecite{KuceraStreda}.
For the fractional regime, it matches the results of
Refs.~\onlinecite{Kivelson} and \onlinecite{ChklovskiLee}, after the
latter are somewhat corrected to allow for the fact that the maxima of
$\sigma_{xx}(B)$ of the integer peaks do not map exactly onto the
maxima of $\sigma_{xx}(B)$ for the principal
series.\cite{ChklovskiRuzin} We can also refer the reader to quantum
Monte-Carlo studies in Ref.\onlinecite{Bhatt} where the same value,
$0.5e^2/h$, for the integer peaks was obtained in a simulation of
single-electron scattering off short-range impurities.  Thus, while
different theoretical models agree on an expected universality of the
conductivity peak heights and even on their values, experiment offers
no evidence to support this prediction.

To make the situation even more confusing, a puzzling feature was
observed in a number of experiments performed at very low temperatures
(15-40 mK): the relative heights of most of the conductivity peaks
obtained in the fractional regime were, indeed, found to scale
approximately as $|\sigma_2-\sigma_1|$, but with an absolute factor
which differed from $1/2$ and, moreover, varied from sample to
sample.\cite{PeakHeightExp} Such a ``universality within one sample'',
while there is none between different samples, is hard to
understand. This feature was especially well seen in recent
experiments performed in the Corbino geometry\cite{RokhimsonGoldman}
where scaling of the peak heights was observed simultaneously for the
fractional and integer regimes, i.e., in quite different ranges of
magnetic field. As a function of temperature, the height of each peak
was found to pass through a maximum value. These maximum values
differed for different peaks, being scattered below, though not very
far from, the universal values predicted by theory.  On the low
temperature side of these maxima, therefore, with decreasing
temperature the peak heights fell further below the universal values,
as has also been earlier observed in low-mobility samples.\cite{Tsui}
At the lowest temperatures (14 mK), however, different peaks in a
given sample converged to the corresponding theoretical values
multiplied by the same sample-dependent factor.

In this work, we suggest an explanation of the phenomenon of the
``non-universal'' scaling of the peak heights. Our explanation assumes
that in the interior of the sample there exist random inhomogeneities
of the electron density with a very large correlation length.  While
the origin of these inhomogeneities is unclear, we note that the
existence of such fluctuations was also a necessary assumption in
recent work\cite{SimonHalperin} which proposed an explanation for
another experimental puzzle -- the Resistivity Law, which is observed
in some samples at higher temperatures.  We can only speculate as to
whether such fluctuations might result from imperfections in the
doping process or from incomplete equilibration of the electron
density on sample cool-down, or from both. The fact that the
experimental traces in the inter-plateau regions often change after
the sample is reheated and cooled down again, indicates that that the
latter mechanism may be important.  By whatever means they might
arise, we argue that even small fluctuations of the electron density
become crucial at low enough temperatures, where the dependence of the
conductivity tensor on the local value of the filling factor is almost
singular.  In particular, we show that, due to the finite size of the
sample, the sample edge gives rise to a positive contribution to the
Corbino resistance which is proportional to
$1/(\sigma_2-\sigma_1)$. This contribution increases with decreasing
temperature due to the growing correlation radius of the random
clusters which are responsible for the current transfer.  At
sufficiently low temperatures, when the correlation radius exceeds the
sample size, the edge contribution dominates the sample
conductance. The heights of the observed peaks in $\sigma_{xx}$ in a
given sample are found to freeze at values which differ from the
corresponding ``universal'' values in an infinite sample by a random
geometric factor which is the same for all peaks.

By making use of a current-voltage duality that exists in
two-dimensional conductivity problems, we show how our results may
also be applied to the Hall bar geometry.  We find that macroscopic
inhomogeneities have the same effect on the measured peaks in
$\rho_{xx}$ in this geometry as on the peaks in $\sigma_{xx}$ for the
Corbino disc. This is in agreement with the fact that sample-dependent
scaling of the peak heights has been observed in both
geometries\cite{PeakHeightExp,RokhimsonGoldman}.

It has previously been realized that density inhomogeneities may
distort conductivity measurements in the Hall bar geometry.  In
particular, for certain gated or mesa-etched systems, the boundary to
a vacuum is believed to involve a rather gradual decrease in electron
density.  This creates a strip along the edge with a quantized Hall
conductivity and very low scattering, while the bulk of sample can be
in a transition region with noticeable dissipation.  Such a strip can
trap a significant portion of the current, making the current
distribution in the sample inhomogeneous and affecting the measured
resistance $R_{xx}$.  In the language of the edge-transport theory,
this can be re-formulated as poor equilibration between the edge
states and the bulk. While this model has been successful in
accounting for certain non-local resistance measurements, the presence
of a smooth edge cannot explain the observed non-universal scaling of
the conductivity peak heights.  Therefore, since the random bulk
inhomogeneities are crucial for this scaling, and since the effects of
a smooth edge have been discussed
before,\cite{McEuen,vanWees,CooperChalker} in this work we will focus
only on the consequences of macroscopic bulk inhomogeneities.  The
results we present for the Hall bar therefore apply to samples with
sharp edges (i.e., with the edge width less than magnetic length).
  For certain samples, some combination of the two models
may be appropriate.

As well as giving rise to ``non-universal scaling'' of the peak
heights at the lowest temperatures, we show that macroscopic
inhomogeneities in the bulk also lead to an enhancement of the
non-local resistance.  In fact, by developing a general
``boundary-strip'' approach which describes finite-size effects in a
macroscopically inhomogeneous sample, we show that, for a sample with
a sharp edge, both of these effects are directly related.  Using this
relation, we propose an experimental method by which the edge
contribution in the observed resistivity may be separated from the
bulk contribution, at least over the range of temperature for which
the edge contribution is not too big (at intermediate $T$).

The paper is organized as follows. In Sec.~II, we discuss the role of
macroscopic inhomogeneities in the transport properties of an infinite
QHE system in different regimes of temperature. Finite-size effects on
the low-temperature two-terminal resistance of an inhomogeneous
Corbino disc are considered in Sec.~III and in the {\it Appendix}.
The boundary impedance matrix formalism, which provides a description
of small edge contributions to the measured resistances in any sample
geometry (Corbino disc, Hall bar etc.), is developed in Sec.~IV. The
explicit form of the impedance matrix is evaluated for a
macroscopically inhomogeneous sample with a sharp edge. The enhanced
``non-local'' resistance arising from the macroscopic inhomogeneities
is studied in Sec. V, where it is also shown that this is directly
related to the edge corrections in the measured ``local'' and Hall
resistances.  Sec. VI concludes the paper.

\section{infinite sample}

We consider an infinite sample in the presence of macroscopic
inhomogeneities of the electron density. The local value of the
filling factor can be written as $\nu({\bbox r})=\overline{\nu}
+\delta\nu({\bf r})$, where $\overline{\nu}$ is the average filling
factor, and $\delta\nu({\bbox r})$ is a small fluctuating component
with a magnitude $\delta\nu_0\ll\overline\nu$. We assume that some
impurity scattering occurs on scales much smaller than $R_c$, such
that the local conductivity tensor exhibits the quantum Hall effect.
Specifically, we will assume that at a critical value $\nu_c$, the
Hall conductivity undergoes a sharp crossover between two quantized
values, $\sigma_1=e^2\nu_1/h$, and $\sigma_2=e^2\nu_2/h$. Here $\nu_1$
and $\nu_2$ are a pair of adjacent values of $\nu$ at which the
electron system is incompressible, $\nu_1<\nu_c<\nu_2$.  The diagonal
conductivity $\sigma_{xx}(\nu)$ has a sharp peak in the crossover
region, being very small everywhere else. The width of the crossover
region $\delta\nu_T$ vanishes as $T\rightarrow 0$.  We can consider
this picture as simply following from numerous experimental data on
the low-temperature quantum Hall effect in large samples.

  The critical behaviour of the conductivity tensor has been
much studied. The width of the crossover region vanishes as
$T\rightarrow 0$ according to
\[ \delta\nu_T\sim T^\kappa, \]
with an exponent $\kappa$ which has been argued to take the universal
value $3/7$.\cite{PolyakovShklovski} In the zero temperature limit,
the maximum value of the dissipative conductivity is thought to
approach $e^2/2h$ in the spin-split integer quantum Hall
effect,\cite{Bhatt} and $(\sigma_2-\sigma_1)/2$ for the fractional
regime.\cite{DykhneRuzin,RuzinFeng} As explained in the introduction,
this prediction was deduced from the hypothesis that the correlated
electron state in the vicinity of a QHE transition represents a random
mixture of two incompressible liquids with localized quasiparticles on
top. The correlation radius $R^{\rm mic}_c$ of this two-phase system
was assumed to be larger than the magnetic length $(\hbar
c/eB)^{1/2}$.  To avoid confusion, we emphasize that the correlation
radius of the macroscopic density fluctuations $R_c$ that we consider
in the present work is assumed to be much larger than $R^{\rm mic}_c$
which is thus considered as a microscopic length. The ``local''
conductivity tensor $\hat\sigma({\bbox r})$ introduced above is
defined at scales larger than $R^{\rm mic}_c$ (and, if effects of
quantum interference are important, larger than the phase-breaking
length) but smaller than $R_c$.  For our purposes, the peak value of
the local dissipative conductivity will not be important; we shall
only assume that it is either of the order of or less than the
difference in the quantized values of the Hall conductivity,
$\sigma_2-\sigma_1$. We believe that this assumption agrees with
existing data on the quantum Hall effect in a strong magnetic field.

Let the magnetic field be tuned so that the average filling factor
$\overline{\nu}$ is close to $\nu_c$. At sufficiently low
temperatures, the crossover width $\delta\nu_T$ becomes much less than
the fluctuation magnitude $\delta\nu_0$ which is small but temperature
independent.  Hence, as temperature goes down, the conductivity
distribution becomes strongly inhomogeneous. In most of the sample, as
illustrated in Fig. 1(a,b), the Hall conductivity $\sigma_{xy}$ is
quantized at either $\sigma_1$ or $\sigma_2$, and the diagonal
conductivity $\sigma_{xx}$ is very small. Only a narrow intermediate
region within the interval $|\nu-\nu_c| \sim \delta\nu_T$, has a
noticeable $\sigma_{xx}$ (this region is indicated by the grey color
in Fig. 1(c)).  As long as $\overline{\nu}$ stays within this
interval, the ``grey'' region forms an infinite percolation
cluster. The latter consists of strips of a small width $w_T\sim
R_c\,\delta\nu_T/\delta\nu_0$ which join near saddle points (critical
vertices).  The characteristic size of a cluster cell, i.e., the
distance between two vertices follows from classical percolation
theory\cite{Isichenko}
\begin{equation}
 \xi_T \sim
R_c(\delta\nu_0/\delta\nu_T)^{4/3}.
\label{xiT}
\end{equation} 
We note that the true geometry of the percolating cluster includes also
loops and by-passes on links between vertices not shown in Fig. 1(c).

The evaluation of the effective conductivity tensor $\sigma^*$
requires an understanding of how the current density is distributed in
such an inhomogeneous system. The current distribution is known to be
quite different in the two limiting cases which are reached as the
relative values of the dimensionless parameters $w_T/R_c$ and
$\sigma_{xx}^{\rm max}/(\sigma_2-\sigma_1)$ are varied.

When $\sigma_{xx}({\bbox r})$ is very small (how small will be
determined below), the continuity conditions force the current to flow
almost exactly along lines of constant
$\sigma_{xy}$.\cite{Ruzin,CooperChalker} For an infinite random
system, this means that the current is concentrated near the
percolation threshold of the function $\sigma_{xy}({\bbox r})$, within
a sparse percolation network inside the grey area shown in
Fig. 1(c). The characteristic parameters of this new network -- the
width $w$ and the length $l$ of an elementary link of the cluster --
are determined by the value of $\sigma_{xx}$ at the percolation level
in $\nu$. They can be estimated from the self-consistent condition
that the current is able to cross the lines of constant $\sigma_{xy}$
in order to pass from one critical saddle point of the network to the
next one, which has a slightly different value of
$\sigma_{xy}$.\cite{Isichenko,SimonHalperin} We give here the final
expressions for both parameters

\begin{eqnarray}
&{\displaystyle
w\sim w_T\left(\frac{\sigma_{xx}}{\sigma_2-\sigma_1}\right)^{3/13}
\left(\frac{R_c}{w_T}\right)^{10/13}},&\\
\label{w}
&{\displaystyle
 l\sim R_c\left(\frac{(\sigma_2-\sigma_1)R_c}{\sigma_{xx}w_T}\right)^{7/13}.}&
\nonumber
\end{eqnarray}
The net diagonal conductivity of the system $\sigma_{xx}^*$ is determined
by the geometry of an elementary link, as given by
$\sigma_{xx}^*\sim \sigma_{xx}{l}/w$ which yields, for the
maximum value of $\sigma_{xx}$

\begin{equation}
\sigma_{xx}^{\rm *\, max}\sim  \left(\sigma_{xx}^{\rm max}
\right)^{3/13}
(\sigma_2-\sigma_1)^{10/13}\left(\frac{\delta\nu_0}{\delta\nu_T}
\right)^{10/13}.
\label{heighthighT}
\end{equation}
Note that the effective conductivity decreases with increasing
temperature, more or less as $1/\delta\nu_T$, unless the temperature
dependence $\sigma_{xx}^{\rm max}(T)$ is very sharp.

The net Hall conductivity $\sigma_{xy}^*$ does not depend on the
geometry of the network and simply coincides, apart from small
corrections, with the percolation threshold value
$\sigma_{xy}(\overline\nu)$.  Hence, $\sigma_{xy}^*$ crosses over from
$\sigma_1$ and $\sigma_2$ within the same interval of $\overline\nu$
as does the local Hall conductivity,
$|\overline\nu-\nu_c|\sim\delta\nu_T$.

This picture is correct provided $\sigma_{xx}$ is sufficiently small
and the grey regions are not too narrow, such that the current can
stay within these regions, $w\ll w_T$.  The last condition can be
written as $T\gg T_{s1}$, with $T_{s1}$ given by the equation
\begin{equation}
\delta\nu_T(T_{s1})=\delta\nu_0
\left(\frac{\sigma_{xx}(T_{s1})}{\sigma_2-\sigma_1}\right)^{3/10}.
\label{Ts1}
\end{equation} 
The regime exists only if $\sigma_{xx}$ is much less than
$\sigma_2-\sigma_1$ at temperatures at which the system can already be
considered to be macroscopically homogeneous, $\delta\nu_T(T)\sim
\delta\nu_0$.

As the temperature is reduced below $T_{s1}$, the current spills out
into the quantum Hall regions. In the limit $T\ll T_{s1}$, which will
be mostly considered in the rest of the paper, the currents flow
predominately in the quantized Hall regions. The details of the grey
areas become unimportant (except near the saddle points), and they may
be replaced by sharp boundaries. This represents a particular example
from a class of ``black-and-white'' systems which was studied in
Refs.\onlinecite{Milton,DykhneRuzin,RuzinFeng}.  Due to the continuity
conditions and the absence of scattering in the bulk, the currents in
the ``black'' and ``white'' phases cannot cross the phase
boundary. Instead, the currents have to focus at vertices to pass
between the two corners of the same color. The net conductivity tensor
depends on the ratio of the average currents flowing in the two phases
which is controlled by $\overline\nu$.  At
$\nu_c-\overline\nu\gg\delta\nu_T$, all saddle points have values of
$\sigma_{xy}$ close to $\sigma_1$ which ensures good percolation in
the ``black'' color, so that white current is relatively small and
$\sigma_{xy}^*\simeq\sigma_1$.  At $\overline\nu-\nu_c\gg\delta\nu_T$,
we get good percolation in ``white'', and
$\sigma_{xy}^*\simeq\sigma_2$.  A crossover takes place within the
interval $\delta\nu_T$ when both currents are comparable. Since the
local diagonal conductivity is small in both quantized regions, the
net conductivity $\sigma_{xx}^*$ is also small when either of these
phases percolates freely, and experiences maximum in the crossover
region.  In the limit of zero dissipation in each of the quantized
Hall regions, the dependence of $\sigma_{xx}^*\ vs. \ \sigma_{xy}^*$
of the isotropic two-phase system is known to be a universal function
-- a semicircle\cite{RuzinFeng}

\begin{equation}
(\sigma_{xx}^*)^2+\left(\sigma_{xy}^*-\frac{\sigma_1+\sigma_2}{2}
\right)^2=
\left(\frac{\sigma_1-\sigma_2}{2}\right)^2.
\label{semic}
\end{equation} 
In particular, the
maximum value of the average conductivity is 

\begin{equation}
\sigma^{\rm *\, max}_{xx} =
(\sigma_2-\sigma_1)/2,
\label{universal}
\end{equation}
independent of the details of the ``grey'' areas.

To conclude, the long-range inhomogeneities in the infinite sample do
not essentially alter the conductivity peak width. They cause the peak
height in $\sigma_{xx}^*$ to saturate at low temperatures at the
universal value (\ref{universal}). If the microscopic conductivity
tensor components $\sigma_{xx}, \sigma_{xy}$, as argued in
Refs.\onlinecite{DykhneRuzin,RuzinFeng}, also satisfy the
``semicircle'' relation (\ref{semic}), then the presence of the
macroscopic inhomogeneities has no effect in the low-temperature
limit.  In a finite sample, however, as we now proceed to show,
$\sigma_{xx}^*$ at $T=0$ deviates drastically from the universal value
due to the inhomogeneities.

\section{Finite-size effects in the Corbino geometry}

To study finite-size effects, we have to resort to a realistic
experimental set-up which means that we have to specify the geometry
of the sample and the attachment of the contacts. Experiments are
performed on the Hall bar, in the van der Pauw method, and on the
Corbino disk. Our initial choice will be the Corbino disc on which
recent experiments in Ref.\onlinecite{RokhimsonGoldman} have been
performed.  In Sec. III, we show how to transfer our results to the
Hall bar using the current-voltage duality. We will not discuss the
van der Pauw geometry explicitly in this work.

\subsection{Contact resistance}

A Corbino disc cut from an inhomogeneous sample is shown schematically
in Fig. 2(a).  The two-terminal resistance between the metal probes
attached at the inner and outer circular edges is measured. We
consider an ideal contact without any tunnel barriers or dielectric
layers between metal and sample, the metal boundary having a constant
potential. We assume, first, that the correlation radius of the
``grey'' cluster $\xi_T$ is much less than all sample dimensions,
$r_1, r_2$ and $W\equiv r_2-r_1$. In the limit $\xi_T/W\rightarrow 0$,
the sample can be considered to be homogeneous with the conductivity
tensor $\hat\sigma^*$. The two-terminal resistance is then given by

\begin{equation}
R_0=\frac{A_0}{\sigma_{xx}^*},
\ \ \ A_0=\frac{1}{2\pi}\ln\frac{r_2}{r_1},
\label{R0}
\end{equation} 
where $A_0$ is the geometric aspect ratio, which for $W\ll r_1,r_2$ is
close to $A_0\simeq W/2\pi r$. A finite value of $\xi_T/W$, as we now
demonstrate, leads to an increase in the two-terminal resistance above
this value (\ref{R0}) due to an effective contact contribution.

The simplest way to estimate this correction is by monitoring the
Joule heat dissipated in the sample. At low temperatures, $T\ll
T_{s1}$, as already discussed, the currents flow in the (black or
white) quantized Hall regions.  Since scattering in the bulk is
negligible, the currents cannot cross the phase boundaries and , in
order to pass through the sample, must focus at vertices formed by
adjacent corners of different phases.  All the energy dissipation in
the system occurs in ``hot spots'' at the vertices.  The vertices are
of two kinds: 4-vertices in the sample interior, formed by alternating
``black'' and ``white'' corners, and 3-vertices at the boundary,
``black-white-metal''. The above resistance (\ref{R0}) corresponds to
dissipation at internal vertices, as given by
 
\begin{equation}
 Q_{\rm int}=I^2R_0.
\label{Qint}
\end{equation} 
Since the metal contact boundary is at a constant potential, the
current lines have to focus at 3-vertices in order to enter the (e.g.,
inner) contact.  This causes additional dissipation at the edge given
by
\[Q_{\rm edg}^{(1)} = \sum_{\alpha} I_\alpha^2 
R_\alpha,\] 
where $I_\alpha$ is the current passing to the metal at the 3-vertex
$\alpha$, and $R_\alpha$ is the effective resistance of the vertex.

It turns out that focusing of the current lines occurs only at those
of the 3-vertices at which the Hall conductivity $\sigma_{xy}$
increases from $\sigma_1$ to $\sigma_2$ to the right when looking into
the sample from the metal (we call these ``active'' vertices). For the
remaining half of the vertices, $I_\alpha=0$. As shown in the {\it
Appendix}, the resistance of an active 3-vertex is always given by
\begin{equation}
R_\alpha=R_3\equiv \frac{1}{2(\sigma_2-\sigma_1)},
\label{R3}
\end{equation} independent of the microscopic details of the vertex core. In
particular, the result remains valid if the phase boundary branches
when approaching the metal forming a fork-like structure, as
illustrated in Fig. 2(b) in magnified view. Branching, though not
shown in Fig. 2(a), does occur in a random percolation cluster on
scales smaller than $\xi_T$. Only the fork vertices with white as the
rightmost color are active.

For a uniform pattern of white/black regions, the current entering the
contact will be shared approximately equally between the $\pi
r_1/\xi_T$ vertices at the edge so that

\begin{equation}
Q_{\rm edg}^{(1)}\sim I^2R_3\frac{\xi_T}{\pi r_1}.
\label{Qedg}
\end{equation}
The total
two-terminal resistance $R_{\rm tot}$ can be found from the total
Joule heat,
\[Q_{\rm tot}=I^2R_{\rm tot}=Q_{\rm edg}^{(1)}
+Q_{\rm edg}^{(2)}+Q_{\rm int}.\] 
Using Eqs. (\ref{R0})-(\ref{Qedg}), we find

\begin{equation}
R_{\rm tot}=\frac{A_0}{\sigma_{xx}^*} + 
\left(\frac{1}{r_1}+\frac{1}{ r_2}\right)\frac{\xi_T}{2\pi
 (\sigma_2-\sigma_1)}.
\label{Rtot}
\end{equation} 
The second term in this expression is, of course, an estimate.
In the next Section, we re-derive Eq. (\ref{Rtot}) rigorously
 for a simple model of a periodic chessboard two-phase
distribution with $\xi_T$ replaced by the square size.

Thus, at the maximum of $\sigma_{xx}^*(\overline\nu)$, the
relative correction to the peak resistance arising from the edges is
of the order of $\xi_T/W$. (We assumed here that $r_1$ is not much less
than $r_2$, and used the fact that, at the lowest temperatures,
$\sigma_{xx}^{\rm *\, max}\sim
\sigma_2-\sigma_1$).  The experimentally measured value of the
diagonal conductivity
$\sigma_{xx}^{\rm exp}$ defined by \cite{RokhimsonGoldman}
 \[\sigma_{xx}^{\rm exp}=A_0/R_{\rm tot}\]
is now lower
than the ``bulk'' conductivity $\sigma_{xx}^*$ by the same relative
amount.  This
negative correction increases as temperature is lowered, since $\xi_T$
becomes larger [Eq. (\ref{xiT})]. 
We believe that the decrease in $\sigma_{xx}^{\rm exp}$ that we
predict from these considerations provides a possible explanation for
the observations of Rokhinson {\it et al.}\cite{RokhimsonGoldman}.

Note that the correction to $\sigma_{xx}^{\rm exp}$ may be different
for different conductivity peaks since the width $\delta\nu_T$ and,
hence, the correlation length $\xi_T$, may vary from peak to peak.
This accounts for the dispersion of traces $\sigma_{xx}^{\rm max}(T)$
observed for different peaks in the integer regime, see data for two samples by
Rokhinson {\it et al.}\cite{RokhimsonGoldman} which we reproduced here in Fig. 3.
  At the lowest temperatures,
however, as one can see from the figure, all the traces converge and tend to 
collapse onto
one curve.  The corresponding low-temperature value can be written as
$\sigma_{xx}^{\rm max}=ke^2/2h$, where the coefficient $k$ is almost
the same for different peaks, but is
sample-dependent\cite{RokhimsonGoldman} (compare Figs. 3(a) and 3(b)).
  As found in the quoted
work, the well-pronounced peaks in the fractional regime have the
height $k(\sigma_2-\sigma_1)/2$, with approximately the same $k$ as do
the peaks obtained on the same sample in the integer regime.  This
agrees with earlier data on the fractional quantum Hall effect
mostly obtained in the Hall bar geometry\cite{PeakHeightExp}
which show that, at the lowest temperatures, most of the peak heights
are proportional to the difference in adjacent plateaus in
$\sigma_{xy}$, with a prefactor that fluctuates from sample to
sample. To understand the origin of the curious ``universality within
one sample'' let us consider the lowest temperatures.

\subsection{Low-temperature limit} 

Since the characteristic correlation radius $\xi_T$ grows as the
temperature is reduced, it will eventually become of the same order as
the distance between the contacts $W$ at some temperature $T\sim
T_{s2}$. At this point, the edge contribution to the resistance in
Eq. (\ref{Rtot}) is as large as the bulk contribution. When
temperature is decreased further, $T\ll T_{s2}$, a strong inequality
$\xi_T\gg W$ is met for any peak, so that no critical 4-vertices
(shown in Fig. 2(a) for higher temperatures) fall within the sample.
The current is transported directly from contact to contact by one or
more pairs of ``white'' and ``black'' clusters connecting the
contacts.  ``Black'' and ``white'' clusters of the size $W$ can exist
simultaneously while $\overline\nu$ is in the interval
$|\overline\nu-\nu_c|\lesssim \delta\nu_w$, where $\delta\nu_w$ is
determined from the equation

\begin{equation}
W\sim R_c\left(\frac{\delta\nu_0}{\delta\nu_w}\right)^{4/3}.
\label{width}
\end{equation}
Outside of this interval, percolation between contacts can exist only
in one phase, either in the ``white'' or in the ``black''.  The resistance
in this case is equal to that of a homogeneous Corbino sample with no
scattering inside, that is, $R_{\rm tot}=\infty$, and $\sigma_{xx}^{\rm
exp}=0$. Hence, $\delta\nu_w$ in Eq. (\ref{width}) represents the
observed peak width.

Notice that $\delta\nu_w$ does not depend on temperature and is much
larger than the value for the infinite sample $\delta\nu_T$. Since no
4-vertices are now found in the sample, the finite value of
$\delta\nu_T$ is no longer relevant. All temperature dependence has
now disappeared from the problem and all the peaks $\sigma_{xx}^{\rm
exp}(\overline\nu)$ must become identical.  We suggest that this
explains why the experimental traces $\sigma_{xx}^{\rm exp}(T)$ in
integer regime converge at low
temperatures\cite{PeakHeightExp,RokhimsonGoldman}. Recall that it was the differing
values of $\delta\nu_T$ that caused the contact contributions to the
total resistance to vary for different peaks at $T\gg T_{s2}$.

This saturation of the peak width which we predict at low temperatures
is similar to the well-known saturation effect which is expected to
occur in small samples when the coherence length $\xi_c$ becomes
comparable to the sample size.\cite{KnownSizeEff}  The observable
difference between the two mechanisms is in the size dependence,
$\delta\nu_w\propto W^{-\kappa}$: we predict the classical index
$\kappa =3/4$ instead of the $3/7$ which is thought to be appropriate
for the quantum problem.\cite{3/7rev,KnownSizeEff}

We now discuss the height and possible shape of the peak.  As we shall
show below, in the zero-temperature limit, the two-terminal conductance
$1/R_{\rm tot}$ can take only quantized values

\begin{equation}
1/R_{\rm tot}=M(\sigma_2-\sigma_1),
\label{quant}
\end{equation} where $M$ is an integer (including zero) which depends on the
specific realization of the disorder. As the average filling fraction
is varied, the value of the integer $M$ may change, and the peak in
$\sigma_{xx}^{\rm exp}(\overline\nu)$ can display an unusual step-like
dependence on filling fraction, as illustrated schematically in
Fig. 4c and 4d.  Although the shape of these peaks (number and
position of the steps) depends on the specific realization of
disorder, this shape is the same, with the peak height expressed in
units of $\sigma_2-\sigma_1$, for all peaks in a given sample.

Let us fix the average filling factor $\overline\nu$ somewhere within
the peak width $\delta\nu_w$, e.g., at the point $\overline\nu=\nu_c$.
We have to distinguish two major cases depending on the random
configuration of a sample.  Either both ``black'' and ``white''
percolate in the radial direction, or both ``black'' and ``white''
percolate in the azimuthal direction, as illustrated in Fig. 4(a) and
4(b). Together with the case in which only one phase percolates in both
directions, which we determined above as being outside of the peak
width, this exhausts all topological possibilities.  In the case of
azimuthal percolation, Fig. 4(b), the current can barely pass between
contacts since it is not allowed to cross the phase boundary. Hence,
peaks are missing, $R_{\rm tot}=\infty$. More precisely,
$\sigma^{\rm exp}_{xx}$ is as small as the average of $\sigma_{xx}$ in the
quantized regions, and therefore rapidly vanishes as $T\rightarrow 0$.
Such behavior is observed in experiment\cite{HongWen}, but it is usually attributed
to a ``bad sample'' or ``bad contacts'' and, consequently,
 does not reach publication.
 In some sense, it is correct
to say this in our model as well; however, both ``bad'' and ``good'' samples belong to 
the
same statistical ensemble with a quite small amplitude of
inhomogeneities.  In the case shown in Fig. 4(a), when the contacts are
connected by one ``black'' and by one ``white'' region the resistance
is finite and equal to the doubled resistance of the 3-vertex $R_{\rm
tot}=2R_3=1/(\sigma_2-\sigma_1)$, see Eq. (\ref{R3}) for the 3-vertex
resistance.

We can easily obtain the shape of the peak in $\sigma_{xx}^{\rm exp}$
for the configuration in Fig. 4(a). Let $\nu_1$ and $\nu_2$ be the filling
factors at the two saddle points which control the current transfer,
$\nu_2-\nu_1 \sim \delta\nu_w$.  When $\overline\nu$ is shifted down
and crosses the level of the lower saddle-point $\nu_1$, the
corresponding ``black'' bridge becomes ``white'', and percolation in
``black'' quits, so that the resistance $R_{\rm tot}$ becomes
infinite. This transition occurs abruptly in $\overline\nu$ (more
precisely, within a small interval $\sim \delta\nu_T$).  Similarly,
for $\bar{\nu}>\nu_2$ the ``white'' region percolates, and the system
is on a quantized Hall plateau corresponding to $\nu_2$.  The
resulting ``peak'' in $\sigma_{xx}^{\rm exp}$ represents a box of a
width $\nu_2-\nu_1$, Fig. 4(c). The latter is of the order of
$\delta\nu_w$ and fluctuates from sample to sample by 100\%.

There may, of course, be more than one pair of ``black'' and ``white''
regions connecting the contacts (and, hence, more saddle points which
switch).  Since all of the 3-vertices at the same edge are connected
in parallel, the inverse resistance is given by Eq. (\ref{quant}),
where $M$ is the number of pairs of connecting regions. In the
degenerate case shown in Fig. 4(b), $M=0$. Thus the total conductance
of the sample is an integer multiple of $\sigma_2-\sigma_1$.  The
number of ``quanta'' $M$ depends on the value of $\overline\nu$. A
typical shape of the peak for $M_{\rm max}=2$ is shown in Fig. 4
(d). Each step in $\sigma_{xx}^{\rm exp}$ results from the switch of
some saddle-point from a ``black'' to a ``white'' bridge.  Note that
the resistance (\ref{quant}) does not demonstrate regular scaling with
the aspect ratio of the sample since $M$ is just a random integer
varying from sample to sample. The probability distribution for
different $M$ does, however, depend on the shape of the sample.  For
instance, if $r_2-r_1\sim r_1$, the most probable values are $M=0, 1$,
or $2$.  In a very narrow ring, $W\ll r$, the most probable values are
close to $M=C/A_0 \gg 1$, where $C$ is some numerical factor. In the
latter case, the resistance will seem to scale properly with the
sample dimensions, as if the sample was homogeneous, except for the
wrong numerical factor.  The quantization of $R_{\rm tot}$ will be
difficult to see.

Our prediction of the {\it longitudinal} resistance quantization
allows a comparison of our theory with the experiments of
Ref.\onlinecite{RokhimsonGoldman}. Although neither of the two samples
shown in Fig. 3 reveals the saturation of the peak heights which
we expect to occur at low enough temperature, still, we can see that
the heights of all peaks  become close to each other at
the lowest temperature studied, $T=14\mbox{mK}$.  Hence we assume that
saturation occurs not very far below this temperature and use the
values of $\sigma_{xx}^{\rm exp}$ obtained at $14\mbox{mK}$ as good
approximations to the zero-temperature values. Taking account of the
aspect ratios quoted in the caption to Fig. 3, for both
samples we obtain $R_{\rm tot}=A_0/\sigma_{xx}^{\rm exp} \simeq h/e^2$
within an accuracy of 10\% (we averaged $\sigma_{xx}^{\rm exp}(T=0)$
over 5 values for different peaks in sample B). This corresponds to $M=1$ in
Eq. (\ref{quant}) which is consistent with our prediction. Thus, at
least for these data, the ``fluctuation'' of $\sigma_{xx}^{\rm exp}$
between samples is simply correlated with the different aspect ratios,
$A_0$, which were used to calculate $\sigma_{xx}^{\rm exp}=A_0/R_{\rm
tot}$.

\section{Boundary strip formalism}
     
So far, we have considered the simplest geometry of the Corbino
disc. Although this method can directly produce the value of
$\sigma_{xx}$ for a homogeneous system, it also has some obvious
disadvantages.  Firstly, it does not allow one to measure the Hall
conductivity. Secondly, it leaves no hope of separating
$\sigma_{xx}^{\rm exp}$ into the contributions arising from the bulk
and those arising from the edge. Both disadvantages stem from the fact
that only one independent experimental parameter (the two-terminal
resistance) is obtained in this method.  To permit a larger number of
independent measurements one must consider another geometry, such as
the Hall bar.  As we will show in this section, macroscopic
inhomogeneities in a finite-sized Hall bar lead to a similar decrease
of the observed peak heights in $\rho_{xx}^{\rm exp}$ as for the
observed peak heights of  $\sigma_{xx}^{\rm exp}$ in the Corbino disc
geometry. In Sec. V, we will suggest a method by which this edge
effect may be compensated in the Hall bar geometry, at least, over the
temperature range for which this contribution is small. At the lowest
temperatures, $T\ll T_{s2}$, as clearly follows from analysis in
Sec. III, all information on the bulk properties is lost beyond
recovery.

For the remainder of this paper we will therefore focus on the regime
of small edge corrections.  In this limit, it is possible to develop a
theory in which the edge effects can be accounted for by a ``boundary
strip''. 
Within this formalism, the Corbino disc and Hall bar are thought of as
homogeneous samples with an infinitesimally thin layer attached at
their boundaries to account for the edge-effects.  This ``boundary
strip'' is characterized by an impedance matrix which linearly relates
the currents in the strip to the potential gradients at the boundary.
Effectively the presence of this strip changes the boundary conditions
on the sample.  Such an approach is valid provided (i) the region
along the edge responsible for the edge effects is narrow and
(ii) the edges are homogeneous along their length. 
The first condition is necessary since otherwise one has to take into account,
in addition to the gradients, 
second and higher derivatives of the electric potential
to describe the boundary impedance. 

In physical devices, there are two main types of edge effects that are
important for transport.  Firstly, the finite-size effects due to the
macroscopic inhomogeneities of the interior of the sample, as
discussed above. This type of inhomogeneity, if present, is equally
important for the Hall bar and the Corbino disc.  Secondly,
smooth-edge effects caused by a gradual change in the electron density
when approaching the sample edge.  One should expect the latter effect
to be much stronger in the Hall bar geometry in which electron density
at the edge is zero.  In principle, the boundary strip formalism is
rather general and can be used to account for both types of
inhomogeneities. For the macroscopic inhomogeneities, both of the
conditions outlined in the previous paragraph are satisfied if the
correlation radius $\xi_T$ is much less then the sample width and, in
the Hall bar, the distance between the voltage probes. Both conditions
would also be satisfied for a Hall bar with a smooth edge if the
voltage probes were small enough as not to affect the edge
properties. In real devices, however, the probes are macroscopically
large and interrupt the edge strip. Within the model of non-local
resistance proposed by McEuen {\it et al.}, based on the edge-state
formalism, the contacts have a strong effect on the current
distribution, since they force equilibration between different edge
channels.\cite{McEuen} Such a model is a clear example of a case in
which the edge is not homogeneous along its length, and for which the
boundary strip formalism does not apply. We will therefore focus only
on cases for which the edge of the Hall bar represents a sharp cut in
the inhomogeneous sample, i.e.  the characteristic width of the edge
region is less than the magnetic length. The boundary properties are
then independent of whether the edge is to vacuum (Hall bar) or to
metal (Corbino disc).  Below we will show how the conductance
properties of such an edge can be related to the parameters of the
two-phase model.  First, we will derive the properties of the boundary
strip for a periodic array of the two regions, and then we will
discuss how these are modified for a random distribution.

As we have seen in Sec. III, the corners formed at each edge by
alternating phases create an effective strip with properties distinct
from that of the interior of the sample.  While discussing the Corbino
disc, the boundary strip has been characterized by a single parameter,
the contact resistance. This is defined as the ratio of the voltage
drop across the strip $V$ to the current $j_\perp L$ crossing the
strip, under the condition that the electric field component along the
edge $E_\parallel$ is zero.  In general, quite different boundary
conditions may be applied. For instance, in the standard Hall bar
measurement, $j_\perp =0, E_\parallel\neq 0.$ We need to develop a
quantitative description for the boundary strip which does not depend
on specific boundary conditions and which therefore applies to all
geometries.  We consider four variables: $V, j_\perp, E_\parallel$,
and the additional current $I$ flowing along the edge in the boundary
strip. The four parameters are related by a contact impedance matrix
$\hat\Sigma$, as given by 
\begin{equation} \left(\begin{array}{l} I\\ j_\perp
\end{array}\right)=
\hat\Sigma \left(\begin{array}{l} E_\parallel\\ V
\end{array}\right),
\label{matrix}
\end{equation} 
which is analogous to the conductivity tensor in the bulk. Due to the
low symmetry of the boundary strip, all four components of
$\hat\Sigma$ are {\it a priori} independent.  The whole inhomogeneous
sample can be thought of as consisting of a homogeneous interior, with
the conductivity tensor of the infinite system $\hat\sigma^*$ and with
the same dimensions as the original sample, and an infinitesimally
thin boundary strip described by the impedance matrix $\hat\Sigma$.

We will now determine the boundary strip impedance matrix for a
simplified model: a periodic two-phase system in which ``black'' and
``white'' are regularly distributed as in a chessboard, and all
vertices have identical scattering properties.   
Since, for a thin boundary strip, the sample shape is not important,
we choose to study a convenient geometry, in which the sample is a
long rectangle with rows of vertices aligned parallel to its
sides. Let the width of the sample be $W=Nd$, where $d$ is the lattice
constant, and $N$ is an integer.  A section of a long sample is shown
in Fig. 5(a).  
Although we are ultimately interested in the case of
large $N$, the periodicity of the problem enables us to use a system
with a small number of vertex rows (in Fig. 5(a), $N=3$).  Our objective
is to replace this system by an equivalent homogeneous sample with
thin boundary strips attached, Fig. 5(b). The new system is
characterized by the electric field $\bbox{E}^\infty$ and the current
density ${\bbox j}^\infty$ in the interior, and the four variables $V,
E_\parallel, I, j_\perp$ describing the boundary strip. The positive
directions for the potential drop $V$ and for the other three
variables are shown by arrows in Fig. 5(b).  All these currents and
fields are assumed to be uniform, representing one particular case
which is sufficient to evaluate matrix (\ref{matrix}).  We will consider
the field ${\bbox E}^\infty$ and the current density ${\bbox
j}^\infty$ to be given as boundary conditions. Even though these two
vectors are related by the conductivity tensor of the infinite system
$\hat\sigma^*$, in what follows, we are not going to use this
relation, and will treat ${\bbox E}^\infty$ and ${\bbox j}^\infty$ as
independent variables.

The criteria for the equivalence of the two systems are as follows.
(i)~The field ${\bbox E}_\infty$ and the current density ${\bbox
j}_\infty$ in the new sample must be the same as the average electric
field and the average current density which would be in the original
sample if the latter was infinite; that is, the properties of the
interior are independent of the presence of the boundaries.  (ii)~The
standard continuity conditions $j_\perp=j_y^\infty,
E_\parallel=E_x^\infty$ must be satisfied.  (iii)~The total current
along the new sample
\begin{equation} 
I_{\rm tot}=2I+Wj^\infty_x
\label{Itotdef}
\end{equation} 
and the total potential drop across the new sample
\begin{equation}
V_{\rm tot}=2V+WE^\infty_y
\label{Vtotdef}
\end{equation}
must be the same as the corresponding quantities in the original system.

To apply these rules, consider the current-field distribution in the
original system, Fig. 5(a). As shown in
Refs.~\onlinecite{DykhneRuzin,RuzinFeng}, each 4-vertex is
characterized by the white-to-white current $I_i$ and the
black-to-black current $J_i$.  Each arrow in Fig. 5(a) denotes a set
of current lines. When passing through the white or black square, the
current spread all over the square, then focuses at a corner where it
can pass to another square. We remind the reader that current lines
cannot cross the boundaries between quantized regions so that the
specific distribution of these lines inside any square is
irrelevant. Each side of a square is at constant potential.  The
potential drop between the two sides forming a (white or black) corner
are given by
\begin{equation}
U_i=\frac{I_i}{\sigma_2}, \ \ V_i=\frac{J_i}{\sigma_1}.
\label{UI}
\end{equation}
where $I_i$ ($J_i$) is the total current focusing at the corner. 
Since the system is periodic, and average fields and currents are
taken to be uniform, the set of local currents $I_i, J_i$ must also be
periodic.  The pair of currents at a vertex takes either of two
values, $I_1, J_1$, for odd vertices, and $I_2, J_2$, for even vertices.
The current crosses a sample boundary by focusing at 3-vertices as
shown in Fig. 5(a) for the case $\sigma_2 >\sigma_1$. This occurs at
every other 3-vertex since current lines can focus only if
$\sigma_{xy}$ experiences a step-like increase to the right when
looking from the edge into the sample (see {\it Appendix}). The splitting of
arrows schematically shows the splitting of the sets
of current lines. A diagram similar to Fig. 5(a) can be drawn for the
 equipotential lines, except the notations $I_i$ and $J_i$ must
everywhere be replaced by $U_i$ and $V_i$ respectively.

The components of the average current density in an infinite sample can be
obtained as an average of the corresponding local currents over vertices
of the two types, as given by
\begin{eqnarray}
j_x^\infty= \frac{I_1-I_2+J_1+J_2}{2d},
\label{jx}\\
j_y^\infty= \frac{I_1+I_2-J_1+J_2}{2d}.
\label{jy}
\end{eqnarray}
Analogously, for the electric field components we have
\begin{eqnarray}
E_x^\infty&=&\frac{1}{2d}\left(-\frac{I_1+I_2}{\sigma_2}+
\frac{J_1-J_2}{\sigma_1}\right),
\label{Ex}\\
E_y^\infty&=&\frac{1}{2d}\left(\frac{I_1-I_2}{\sigma_2}+
\frac{J_1+J_2}{\sigma_1}\right),
\label{Ey}
\end{eqnarray}
where we used Eqs. (\ref{UI}). The total current along the sample
$I_{\rm tot}$ can be evaluated by adding the local currents crossing
the vertical dashed line shown in Fig. 5(a). The result can be written
as
\begin{eqnarray}
&{\displaystyle I_{\rm tot}= Nd\,j^\infty_x+2I},&
\label{Itot}\\
&{\displaystyle I=\frac{1}{4}\left(I_1+I_2+J_1-J_2\right)}&,
\label{I}
\end{eqnarray}
where $j^\infty_x$ is given by Eq. (\ref{jx}).
The total potential drop across the sample works out to be

\begin{eqnarray}
&{\displaystyle V_{\rm tot}= Nd\,E^\infty_y+2V,}&
\label{Vtot}\\
&{\displaystyle V=\frac{1}{4}\left(
\frac{I_1+I_2}{\sigma_2}+\frac{J_1-J_2}{\sigma_1}\right)}&,
\label{V}
\end{eqnarray}
where $E^\infty_y$ is given by Eq. (\ref{Ey}).  As we can see from
condition (iii) [Eqs. (\ref{Itotdef}), (\ref{Vtotdef})] the parameters
$I$ and $V$ introduced in Eqs. (\ref{I}) and (\ref{V}) represent, by
definition, the effective current in the strip and the effective
voltage drop at the strip, respectively.  Notice that the right-hand
sides of Eqs. (\ref{I}), (\ref{V}) depend only on $I_1+I_2$ and
$J_1-J_2$. Using the system of Eqs. (\ref{jy}), (\ref{Ex}), we can
express these two linear combinations in terms of components
$j_y^\infty, E_x^\infty$. On the other hand, from condition (ii) we
have $j_y^\infty=j_\perp$ and $E_x^\infty =E_\parallel$. As a result,
we arrive at two equations relating $I$ and $V$ to $j_\perp$ and
$E_\parallel$. This can be written in the form (\ref{matrix}) with
matrix $\hat\Sigma$ given by
\begin{equation}
{\displaystyle \hat\Sigma=\left(
\begin{array}{ll}
{\displaystyle -\frac{d}{4}(\sigma_2-\sigma_1)\ \ }
 &  {\displaystyle\frac{1}{2}(\sigma_2+\sigma_1)}\\
 \ &\  \\
  {\displaystyle -\frac{1}{2}(\sigma_2+\sigma_1)\ \ }
 & {\displaystyle\frac{1}{d}(\sigma_2-\sigma_1)}
\end{array}
\right)}
\label{matrixfin}
\end{equation}
Note that during this derivation we did not make any particular
assumptions about the conductivity tensor in the bulk $\hat\sigma^*$
since we did not specify components of ${\bbox j}^\infty$ and ${\bbox
E}^\infty$.

Let us discuss possible modifications of our result for the case in
which the two-phase system is random with a correlation radius
$\xi_T$, Eq. (\ref{xiT}). Obviously, the Hall components $\Sigma_{12}$
and $\Sigma_{21}$ in Eq. (\ref{matrixfin}) will not change since they
simply reflect the fact that black and white phases, in the vicinity
of the inter-plateau crossover, share the edge equally. Then, the
diagonal components $\Sigma_{11}$, $\Sigma_{22}$ have to be
proportional to $\sigma_2-\sigma_1$ since they originate from the
dissipative resistance of 3-vertices, Eq. (\ref{R3}). The lattice
period $d$ in the two components $\Sigma_{11}$, $\Sigma_{22}$ has to
be replaced by some average lengths, $d_1$ and $d_2$, respectively,
both of the order of $\xi_T$. One might expect {\it a priori} that
$d_1$ and $d_2$ could differ by a numerical factor, since directions
along and across the edge are not equivalent. However, as we show
below, both lengths are exactly equal, $d_1=d_2\equiv d$.

We will employ a duality that exists between the current and field
distributions in 2D conductors.\cite{Dykhne} Let us imagine that the
system in Fig. 5(a) (with many rows of vertices) is randomized, i.e.,
squares are distorted and vertices are not identical.  The currents
$I_i, J_i$ and the voltage drops $U_i, V_i$ at corners are no longer
periodic. Although the system is fully characterized by the discrete
set of currents and voltages at the vertices, it will be convenient
for now to consider the local current density distribution ${\bbox
j}({\bbox r})$ and the local electric field ${\bbox E}({\bbox
r})$. Both functions satisfy the continuity conditions that the number
of current or potential lines entering and leaving a given (black or
white) square are equal.  Suppose that we have found the contact
impedance matrix in this system which, as explained above, has a form

\begin{equation}
{\displaystyle\hat\Sigma=\left(
\begin{array}{ll}
{\displaystyle -\frac{d_1}{4}(\sigma_2-\sigma_1)\ \ }
 &  {\displaystyle\frac{1}{2}(\sigma_2+\sigma_1)}\\
 \ &\  \\
  {\displaystyle -\frac{1}{2}(\sigma_2+\sigma_1)\ \ }
 & {\displaystyle\frac{1}{d_2}(\sigma_2-\sigma_1)}
\end{array}
\right)}
\label{matrixor}
\end{equation} 
Let us now map our system onto a new (primed) system with the same
geometry of the phase distribution and with the new current density
and electric field

\begin{equation}
{\bbox E'}({\bbox r})=[\hat z\times {\bbox j}({\bbox r})],\ \ 
{\bbox j'}({\bbox r})=[\hat z\times {\bbox E}({\bbox r})]
\label{mapping}
\end{equation} 
The steady-state conditions, $\nabla \cdot\bbox{j'}= 0,
\nabla\times\bbox{E'}=0,$ are obviously satisfied in the new system,
if they are satisfied in the original system, $\nabla \cdot\bbox{j}=
0, \nabla\times\bbox{E}=0$. As follows from Eqs.~(\ref{mapping}), the
new quantized Hall conductivities in the black and white are

\begin{equation}
\sigma_k'=-\rho_k\equiv -1/\sigma_k,\ \ k=1, 2.
\label{rho}
\end{equation}
The average components $I, V prior to
publication, j_\perp, E_\parallel$ characterizing
the boundary strip are transformed in the same way as the components
of the local current and electric field in Eqs. (\ref{mapping}), that
is

\begin{equation}
\begin{array}{ll}
V'=I, &E_\parallel '=-j_\perp,\\
I'=-V, &j_\perp '=E_\parallel.
\end{array}
\label{edgemap} 
\end{equation}
Since the geometry of the phase distribution in the primed system does
not differ from the original one, the variables $I', j_\perp '$ should
be related to $E_\parallel ', V'$ via the same impedance matrix
$\hat\Sigma $ given in Eq. (\ref{matrixor}), except $\sigma_k$ should
now be replaced by $\sigma_k'$. Using Eqs. (\ref{rho}),
(\ref{edgemap}), we arrive at the relation

\begin{eqnarray}
&\left(\begin{array}{l}
E_\parallel\\ V
\end{array}\right)=
\hat {\rm P}
\left(\begin{array}{l}
I\\ j_\perp
\end{array}\right),&
\nonumber\\
 &{\displaystyle\hat {\rm P} =\left(
\begin{array}{ll}
{\displaystyle\frac{1}{d_2}(\rho_1-\rho_2)\ \ }
 & {\displaystyle -\frac{1}{2}(\rho_1+\rho_2)}\\
 & \\
  {\displaystyle\frac{1}{2}(\rho_1+\rho_2)\ \ }
 &  {\displaystyle -\frac{d_1}{4}(\rho_1-\rho_2)}
\end{array}
\right).}&
\label{Rho}
\end{eqnarray}
Comparing Eqs. (\ref{matrix}) and (\ref{Rho}), we see that $\hat {\rm
P}=\hat\Sigma^{-1}$. As one can easily check, this is only consistent
with Eq. (\ref{matrixor}) if $d_1=d_2$, which proves our assertion
that $d_1=d_2\equiv d$ even for a random system.

Thus the presence of the edge is equivalent to a fictitious
homogeneous anisotropic strip of width $d/2\sim \xi_T$ with the local
resistivity tensor components
\begin{eqnarray}
&{\displaystyle
\rho_{yx}^b=-\rho_{xy}^b=(\rho_1+\rho_2)/2,}&
\label{rhoyxc}\\
&{\displaystyle
\rho_{xx}^b=-\rho_{yy}^b=(\rho_1-\rho_2)/2.}&
\label{rhoxxc}
\end{eqnarray} 
The unusual fact that the dissipative resistivity in the direction
perpendicular to the strip is negative deserves comment.  A real
physical strip (or a layer if in 3D) with well-defined geometric
boundaries cannot have a negative net diagonal resistivity in any
direction, since this would contradict the Second Law of
thermodynamics.  However, this is not the case here: the effective
contact strip in our discussion has no real geometric boundary which
could be drawn, for instance, inside of the sample in Fig. 4(a).  The
strip describes small corrections to the net conducting properties of
the sample which, as a whole, has a positive dissipation.

We can now reobtain the contact resistance in the Corbino geometry,
say, that from the inner contact $\Delta R_1$. Putting $E_\parallel
=0$, from Eqs. (\ref{matrix}), (\ref{matrixfin}) we have

\begin{equation}
\Delta R_1=\frac{V}{j_\perp 2\pi r_1}=
\frac{d}{2\pi r_1(\sigma_2-\sigma_1)},
\label{reobtain}
\end{equation}
which coincides with the corresponding term in $R_{\rm tot}$,
Eq. (\ref{Rtot}), if one puts $d=\xi_T$.

Consider now the standard Hall bar measurement in which the current
flows parallel to the edges, $j_\perp=j^\infty_y=0$. Relation
(\ref{Rho}) taken with $d_1=d_2=d$ yields
\begin{equation}
E_\parallel =\frac{2\rho_{xx}^b}{d}I,\ \ V=\rho_{yx}^bI,
\label{EV}
\end{equation}
where $\rho_{xx}^b$ and $\rho_{yx}^b$ are given by Eqs. (\ref{rhoxxc})
and (\ref{rhoyxc}).  The electric field in the interior is homogeneous
and given by

\begin{equation}
E^\infty_x=E_\parallel=\rho_{xx}^*j^\infty_x,\ \ 
E^\infty_y=\rho_{yx}^*j^\infty_x.
\label{EE}
\end{equation}
where $\hat\rho^*=\left(\hat\sigma^*\right)^{-1}$ is the resistivity
tensor of an infinite sample.  Expressions for the experimentally
measured components of the resistivity tensor $\rho_{xx}^{\rm
exp}=WE_\parallel/I_{\rm tot}, \rho_{yx}^{\rm exp}= V_{\rm tot}/I_{\rm
tot}$ can easily be obtained from Eqs. (\ref{Itotdef}),
(\ref{Vtotdef}), (\ref{EV}), and (\ref{EE}).  The result has a form

\begin{eqnarray}
&{\displaystyle
 \frac{1}{\rho_{xx}^{\rm exp}}=
\frac{I_{\rm tot}}{WE_\parallel}=
\frac{1}{\rho_{xx}^*}+\frac{d}{W\rho_{xx}^b},}& 
\label{rhoxx}\\
&{\displaystyle
\rho_{yx}^{\rm exp}=
\frac{V_{\rm tot}}{I_{\rm tot}}=
\rho_{yx}^*-\frac{d\rho_{xx}^*}{W\rho_{xx}^b}
\left(\rho_{yx}^b-\rho_{yx}^*\right).} &
\label{rhoyx}
\end{eqnarray} 
The last expression was expanded in terms of the small parameter
$d/W$.

Thus, the presence of the edge leads to a negative correction in the
measured $\rho_{xx}^{\rm exp}$, as it did for $\sigma_{xx}^{\rm exp}$
in the Corbino geometry.  Moreover, if $\rho_1-\rho_2\ll \rho_1$, the
relative magnitudes of both corrections are the same provided the
Corbino disc is narrow and has the same ratio $d/W$ as the Hall
bar. As we have seen above, this fact is related to the current-field
duality.  Physically, the negative correction to $\sigma_{xx}^{\rm
exp}$ in the Corbino measurement results from an additional voltage
drop at the edge, and the negative correction to $\rho_{xx}^{\rm exp}$
in a Hall bar measurement results from an additional current trapped
at the edge. Our conclusion agrees with the experimental observation
that the peak heights tend to decrease at low temperatures in both
geometries.\cite{RokhimsonGoldman,Tsui} The correction to the measured
Hall conductivity changes sign in the middle of the crossover region
where $\rho_{yx}=(\rho_1+\rho_2)/2$. As a result, the characteristic
width of the transition for both functions $\rho_{yx}^{\rm
exp}(\overline\nu)$ and $\rho_{xx}^{\rm exp}(\overline\nu)$ is
somewhat increased by a relative factor of $d/W$.

When calculating the impedance matrix above, we considered the
particular case in which the average current density and electric
field in the interior of the equivalent sample shown in Fig. 5(b) are
homogeneous.  In the following section we will discuss ``non-local''
resistance measurements for which this is not the case.  We now
discuss the applicability of the boundary strip approach for an
inhomogeneous electric field.  Suppose that the electric field in the
interior of the equivalent sample ${\bbox E}^\infty({\bbox r})$ varies
with some characteristic length $l_E$, where $l_E\gg d$. Then the four
parameters of the boundary strip entering matrix relation
(\ref{matrix}) will also depend on the coordinate along the edge
$x$. Let us express the current density in the homogeneous interior in
terms of the pseudoscalar $\psi({\bbox r})$ as given by

\begin{equation}
{\bbox j}({\bbox r})=[\hat{\bbox z}\times\nabla\psi({\bbox r})],
\label{psi}
\end{equation}
which is always possible since $\nabla\cdot{\bbox j}=0$. Analogously,
the electric field can be written in terms of the electric potential
as ${\bbox E}^\infty({\bbox r})=-\nabla \phi({\bbox r})$. At the
boundary strip, $\psi$ experiences a step from $\psi_1$ to
$\psi_2$. The current inside the strip $I$ and the current crossing
the strip $j_\perp$ are given by

\begin{equation}
I=\psi_1-\psi_2,\ \ j_\perp = \frac{1}{2}
\left(\frac{d\psi_1}{dx}+\frac{d\psi_2}{dx}\right).
\label{Ij}
\end{equation}
Analogously, $V$ and $E_\parallel$ are given by

\begin{equation}
V=\phi_1-\phi_2,\ \ E_\parallel = -\frac{1}{2}
\left(\frac{d\phi_1}{dx}+\frac{d\phi_2}{dx}\right).
\label{VE}
\end{equation} 
where $\phi_1$ and $\phi_2$ are the potentials at the outer and inner
sides of the strip, respectively. Thus, the current across the strip
is defined as the average of the currents crossing the inner and the
outer sides of the strip. Since the current inside the strip $I$ depends,
in the general case, on the coordinate $x$, the two currents may be
different. Similarly, $E_\parallel$ is the average of the parallel
components of the electric field at the two sides of the strip. Such a
choice of the definitions of $E_\parallel$ and $j_\perp$ ensures that
all parameters entering the relation (\ref{matrix}) are expressed via
first derivatives of functions $\psi$ or $\phi$: $I$ and $V$ are
discrete derivatives (differences) in $y$, and $j_\perp$ and
$E_\parallel$ are continuous derivatives in $x$. As a result, the
matrix relation (\ref{matrix}) represents the correct description of the
conducting properties of the edge to first order in the small
parameter $d/l_E$. To increase the accuracy to second order, one
would have to write a matrix relation which also includes second
derivatives of $\psi$ and $\phi$ such as $d\psi_1/dx-d\psi_2/dx,
d^2\psi_1/dx^2+d^2\psi_2/dx^2$ etc. In this and the following
sections, we restrict ourselves to first-order effects in
$d/l_E$. We note that the aforementioned equivalence between the
effective boundary strip and a homogeneous strip with a resistivity
tensor $\hat\rho^b$ is also only correct to first-order in this
parameter.

\section{Compensating the finite size effect: ``non-local'' resistance}

The objective of a standard QHE transport experiment performed on a
large sample is to extract the bulk conductivity tensor characterizing
an infinite system.  As we have argued above, at low temperatures
comparable to $T_{s2}$, finite-size effects become noticeable. It
would be very useful to devise a method by which these edge
contributions could be separated from the measured resistances.  In
this section we show how this may be achieved in systems for which the
boundary-strip formalism of section IV applies: that is, for samples
with a sharp edge and for which the edge effects are not too large.

As shown in the previous section, at temperatures which are not too
low, the edge effect can be described by a matrix $\hat\Sigma$ (or
$\hat {\rm P}=\hat\Sigma^{-1}$) which, for the two-phase model with a
sharp edge, contains a single unknown parameter -- the average length
$d\sim \xi_T$.  In order to determine this parameter, one additional
measurement beyond the standard measurements of $R_{xx}$ and $R_{xy}$
in the Hall bar geometry is required.  In what follows, we suggest a
way in which $d$ may be extracted from a measurement of the enhanced
``non-local'' resistance.\cite{McEuen,Goldman}

Unlike the standard Hall bar measurement, for which the current passes
along the sample, in a ``non-local'' measurement the current is forced
to cross the sample between probes 1 and 2 on opposite long sides of
the sample, Fig. 5(b). The ``non-local'' resistance is determined from
the potential difference between a second pair of probes 3 and 4,
$R_{\rm nloc}=V_{34}/I_{12}$.  In a homogeneous sample, as follows
from standard electrostatic considerations, $R_{\rm nloc}$ should
decay with the distance between the current and voltage probes, $L$,
as given by the series
\begin{equation}
R_{\rm nloc}=  \rho_{xx}^*\left(C_1e^{-\pi L/W}+C_2e^{-3\pi L/W}+...
\right),
\label{hom}
\end{equation} 
where $C_1\sim C_2\sim 1$ are numerical coefficients determined by the
shape of the contacts. At $L\gtrsim W$, the resistance is dominated by
the first exponential in this series.  As found in
Ref.\onlinecite{Goldman}, the experimental value of $R_{\rm nloc}$
observed at low temperatures is much larger than that predicted by
Eq. (\ref{hom}). This effect clearly indicates the existence of
currents localized near the edge, in addition to the current passing
in the interior of the sample.
 
The most common explanation for the enhanced ``non-local'' effect
invokes the presence of a smooth edge to the sample, at which the
electron density vanishes slowly.  This would lead to the appearance
of one or more quantized Hall strip(s) at the edge, within which the
scattering can be very small, even when the bulk of the sample is in
the region of the peak in $\rho_{xx}$.  The low-dissipative strip can
trap a noticeable portion of the current in the
sample.\cite{DykhneRuzin} In the edge-state transport language, this
can be formulated as a poor equilibration between different edge
channels: those at the edge of the sample and those in the
bulk.\cite{McEuen,FengNonloc} It is also possible for an enhanced
``non-local'' effect to arise, even when the edge is abrupt, as a
result of the random macroscopic inhomogeneities which we have
discussed in previous sections.  As we have seen above, the effective
boundary strip traps an additional current $I$ along the edge which
causes a decrease in the observed $R_{xx}$.  As we will show in this
section, the same edge current causes an enhanced ``non-local'' effect
which can be thought of as an effective increase in the sample width.

We present a simple quantitative theory which allows one to relate the
corrections in $R_{xx}$ to the enhancement in $R_{\rm nloc}$.  Since
our derivation is based on the phenomenological boundary strip
description, this approach is rather general and can be used for a
class of the edge models.  Moreover, although the edge properties are
described by the four components of the matrix $\hat\rho^b$ which are
expected to depend on a specific edge model, the relation between the
corrections in $R_{xx}$ and the enhancement in $R_{\rm nloc}$ turns
out to be universal in the sense that it does not include any of these
components. The edge corrections to the measured Hall resistance
$R_{yx}$ can be found from the non-local effect in the same manner.
In this case, however, one has to know the component $\rho_{yx}^b$ of
the boundary strip matrix.

We begin by deriving the current and potential distributions for a
non-local measurement on the effective sample shown in Fig. 5(b).  To
do so, it is convenient to express the current density in terms of a
pseudoscalar $\psi({\bbox r})$, as defined by
Eq. (\ref{psi}). Correspondingly, the currents inside and across the
strip are given by Eq. (\ref{Ij}).  The distribution $\psi({\bbox r})$
in the interior of the sample satisfies the Laplace equation

\begin{equation}
\nabla^2\psi=0 ,
\label{nabla}
\end{equation} 
which follows from the conditions $\nabla\times{\bbox E}=0,
 \rho^*_{xx}\neq 0$.  In addition, the parallel component of the
 electric field in the interior $E_x$ and the electric field in the
 strip $E_\parallel$ must match, as given by

\begin{eqnarray}
&E_x(y= -W/2)=E_\parallel,&
\nonumber\\
&E_x=\rho_{xx}^*j_x-\rho_{yx}^*j_y,&
\nonumber\\
&E_\parallel=\rho_{xx}^b2I/d-\rho_{yx}^bj_\perp,&
\label{match}
\end{eqnarray} 
where the matrix $\hat\rho^b$ is defined in Eqs. (\ref{rhoyxc}),
(\ref{rhoxxc}). 
Using Eqs. (\ref{psi}), (\ref{Ij}), the last condition can be written
as
\begin{eqnarray}
&{\displaystyle\frac{2}{d}\rho_{xx}^b(\psi_1-\psi_2)-\frac{\rho_{yx}^b}{2}
\frac{d\psi_1}{dx}+\left(\rho_{yx}^*-\frac{\rho_{yx}^b}{2}\right)
\frac{\partial\psi_2}{\partial x}}& \nonumber\\ &{\displaystyle
+\rho_{xx}^*\frac{\partial\psi_2}{\partial y} = 0,}&
\label{cont}
\end{eqnarray}
where all partial derivatives are evaluated at the lower edge in Fig. 5(b),
$y=-W/2$. The matching condition for the upper edge is analogous.

The solution of Eqs. (\ref{nabla}), (\ref{cont}) depends on
$\psi_1(x)$ , i.e., on the conditions on the outer boundary of the
sample which are set in the experiment. Let us assume that, in the
``non-local'' resistance measurements, the current $I_{12}$ enters and
leaves the sample at small contacts 1 and 2 which are positioned at
$x=0$, Fig. 5(a). Then we have $\psi_1=I_{12}/2$ at $x>0$ and
$\psi_1=-I_{12}/2$ at $x<0$, with a step at $x=0$.  We will restrict
ourselves to finding $\psi({\bbox r})$ far from the current probes,
$x\gg W$. In the limit $x\rightarrow \infty$, all currents vanish so
that $\psi_2\rightarrow\psi_1$. We will look for an asymptotic
solution of Eqs. (\ref{nabla}), (\ref{cont}) in the form
\begin{equation}
\psi({\bbox r})=I_{12}/2+C\cos(ky)e^{-kx}
\label{anzatz}
\end{equation} 
which is symmetric about $y=0$  and
satisfies Eq. (\ref{nabla}).  Substituting this anzatz and
$d\psi_1/dx=0$ into (\ref{cont}), we obtain the following equation
for the decrement $k$
\begin{equation}
\tan\left(\frac{kW}{2}\right)=
\frac{2\rho_{xx}^b}{kd\rho_{xx}^*}+\frac{\rho_{yx}^*-\rho_{yx}^b/2}
{\rho_{xx}^*}.
\label{keq}
\end{equation}

We have to choose the smallest solution of this last equation which
tends to to $\pi/W$ as $d/W\rightarrow 0$. To first order in $d/W$, we
obtain
\begin{equation}
k=\frac{\pi}{W_{\rm eff}},\ \ W_{\rm eff}
=W+d\frac{\rho_{xx}^*}{\rho_{xx}^b}.
\label{k}
\end{equation} 

Clearly, the voltage between probes 3, 4 decays the distance $L$ in
the same exponential way as the current density, such that $R_{\rm
nloc}\propto \exp(-\pi L/W_{\rm eff})$.  Thus, as far as non-local
resistance measurements are concerned, macroscopic inhomogeneities
make the sample effectively wider.  In the region of the peak in
$\rho_{xx}^*(\overline\nu)$ where $\rho_{xx}^*\sim \rho_{xx}^b$, the
effective width increases by of order the correlation radius $\xi_T$.

Notice that the formula (\ref{rhoxx}) for the measured diagonal
resistivity can now be rewritten as
\begin{equation}
\rho_{xx}^*=\rho_{xx}^{\rm exp}\frac{W_{\rm eff}}{W},
\label{xxW}
\end{equation} 
which means that effect of inhomogeneities on the ``local'' resistance
can be thought of as the same increase in the sample width:
$R_{xx}=\rho_{xx}^{\rm exp}L/W=\rho_{xx}^*L/W_{\rm eff}$. This immediately
gives a simple method by which the edge contribution to
$\hat{\rho}^{\rm exp}$ can be eliminated to first order in $d/W$.
First
one must measure the
``non-local'' resistance $R_{\rm nloc}$ with two pairs of voltage
probes situated at different distances $L_1, L_2$ (larger than $W$)
from the current probes. The effective width can be calculated as

\begin{equation}
W_{\rm eff}=\pi(L_2-L_1)\ln^{-1}\left[\frac{R_{\rm nloc}(L_1)}
{R_{\rm nloc}(L_2)}\right].
\label{Weff}
\end{equation}
Combined with a knowledge of the geometric sample width $W$ this can
then be substituted in Eq. (\ref{xxW}) to obtain $\rho_{xx}^*$.
(Since precise measurements of the geometric width of the sample $W$
may present some difficulties in experiment one may determine it in
the same way as $W_{\rm eff}$, but at $B=0$ when weak inhomogeneities
are not important.) Using the same method, we can extract the bulk
value of the Hall resistivity $\rho_{yx}^*$ from observed Hall
resistance $R_{yx}=\rho_{yx}^{\rm exp}$.  From Eqs. (\ref{k}),
(\ref{rhoyx}) we get
\begin{equation}
\rho_{yx}^*=\rho_{yx}^{\rm exp}+(\rho_{yx}^{\rm exp}-\rho_{yx}^b)
\left(\frac{W_{\rm eff}}{W}-1\right).
\label{yxW}
\end{equation}

It is worth emphasizing that when deriving expressions
(\ref{xxW})-(\ref{yxW}), we did not use any particular values for the
components of the boundary strip matrix $\hat\rho^b$. This means that
the method of compensating the edge effect we just have suggested is
rather general and is not restricted to the ``black-and-white''
regime, $T\ll T_{s1}$, studied in Secs. III and IV.  For instance, the
method can be applied just as well at higher temperatures, $T\gtrsim
T_{s1}$, when a significant part of the current is carried by the
intermediate ``grey'' region in Fig. 1(c). This seems to be the case
in experimental data shown  in
Fig. 3 at high temperatures when $\sigma_{xx}^{\rm exp}$ decreases
with increasing $T$.  Relation (\ref{xxW}) used to compensate the edge
effect in the diagonal resistivity is quite universal in the sense
that it does not depend on any components of the boundary strip
matrix.  In order to exclude the edge effect from the Hall resistivity
using Eq. (\ref{yxW}), one has to know the specific value of
$\rho_{yx}^b$.  We note, however, that the value
$\rho_{yx}^b=(\rho_1+\rho_2)/2$ is more general than its derivation
based on the ``black-and-white'' model which we gave in Sec. IV. It
merely reflects the symmetry of the local Hall resistivity
distribution $\rho_{yx}({\bbox r})$ with respect to
$(\rho_1+\rho_2)/2$ which is conserved as long as the sample is
effectively very inhomogeneous, i.e. $\delta\nu_T\ll \delta\nu_0$.

We emphasize again that this method applies only to first order in
$d/W$.  It yields a partial compensation of the edge effect at
$T\gtrsim T_{s2}$, and is useless at $T\ll T_{s2}$ when all measured
parameters have already saturated in temperature.

Our conclusions are easy to test experimentally. If the underlying
model is appropriate in a particular sample, the application of this
method should produce a wide range in the temperature dependence of
$\sigma_{xx}^{\rm max}(T)$ within which the diagrams $\sigma_{xx}$
vs. $\sigma_{xy}$ are close to the ``universal semicircle''. In other
words, by compensating the edge effects, the maxima in temperature
like those shown in Fig. 3  can
be broadened and brought much closer to $0.5e^2/h$. At the lowest
temperatures, when the finite-size effects dominate the transport
properties (the ``saturation'' regime), the method eventually breaks
down and no universality can be retrieved. In fact, such a behavior
close to the universal prediction was already observed, without any
special methods, in the integer regime in the temperature ranges
$T=(0.5-2)\mbox{K}$ [Ref.\onlinecite{Tsui}(a)], and $T=(2-4)\mbox{K}$
[Ref.\onlinecite{HongWen}]. In these low-mobility samples, the
high-temperature decrease of the peak heights starts late when the
size effects are already small.

\section{Conclusions}

We have shown that low-temperature measurements of the quantum Hall
effect in inter-plateau regions are very sensitive to even weak
macroscopic density inhomogeneities in the sample.  The
inhomogeneities may result in strong finite-size effects even in
samples which, from a conventional point of view, are very large.
This may be a reason why the predicted universal behavior of the
transition regions in an infinite macroscopically homogeneous system
is so hard to observe experimentally.  Within our model of macroscopic
inhomogeneities, we were able to account for both the decrease of the
conductivity peak heights at low temperatures and the curious
``non-universal'' scaling of the peak heights revealed in some
samples. In the low-temperature limit, the experimentally observed
peaks in $\sigma_{xx}$ in the Corbino geometry are shown to saturate
at values proportional to the differences between adjacent plateaus in
$\sigma_{xy}$. (Analogously, the peaks in $\rho_{xx}$ observed in the
Hall bar should be proportional to differences in $\rho_{xy}$.)  The
proportionality factor depends on the specific realization of disorder
and fluctuates strongly between samples. The model also predicts
unusual peak shapes which show quantized plateaus in the {\it
longitudinal} conductivity. The experimental peak heights obtained in
Ref.\onlinecite{RokhimsonGoldman} at the lowest temperatures are
consistent with this quantization.

Finally, we also showed that for a Hall bar with a sharp edge, there
exist simple relations between the enhanced non-local resistance and
the size corrections in $R_{xx}$ and $R_{xy}$ when these corrections
are small. These relations can be used to separate the edge effects
from the bulk tensor components $\rho_{xx}^*$ and $\rho_{xy}^*$.

\acknowledgements{ We are thankful to C.W.J. Beenakker,
D.B. Chklovskii, V.J.Goldman, L.P. Rokhinson, and S.H. Simon for
stimulating discussions. We appreciate help from V.J. Goldman,
L.P. Rokhinson, and B. Su who provided us with unpublished experimental data.
 One of us (I.M.R.) expresses his gratitude to the Lorentz
Institute in Leiden, Netherlands, for hospitality. This work was
partially supported by DMR Grant DMR-94-16910, DOE Grant
DE-FG03-88ER45378, ONR Grant N00014-90-J-1829, and by the NATO Science
Fellowship Programme.}

\appendix
\section*{\\Effective Resistance of a Fork-Vertex}

In this Appendix, we calculate the effective resistance $R_3$ of a
fork vertex, a relatively simple example of which is shown in
Fig. 2b. The current distribution in the vertex is shown by the arrows
which denote beams of current lines. The beams split and focus at
``simple'' 3- and 4-vertices which form the fork vertex.  (Recall
that, since the longitudinal conductivity is assumed to be vanishingly
small in black and white regions, the current lines cannot cross the
border between phases or the metal boundary other then at these simple
vertices.) The effective resistance $R_3$ can be found from the net
Joule heat $Q=R_3I^2$, where $I$ is the net current entering the metal
contact. Although the problem would appear quite complex, it turns out
that the dissipated heat does not depend on the specific structure of
the vertex.  We will now show this from very general arguments.

Let us draw an imaginary circle (the dashed line in Fig. 2b) enclosing
all of the structure of the vertex. Consider the total currents $I_1,
I_2$, and $I$ crossing this circle in the black, white, and metallic
areas, respectively. From the current continuity condition,
$I=I_2-I_1$, and from the condition that electric potentials at points
1, 3 of the metal boundary must be equal, $I_1/\sigma_1=I_2/\sigma_2$,
we find
\begin{equation}
I_1=\frac{\sigma_1}{\sigma_2-\sigma_1}I,
\ \  I_2=\frac{\sigma_2}{\sigma_2-\sigma_1}I.
\label{I1I2}
\end{equation}
that is, only one of the three currents is an independent quantity.

The total Joule heat inside of the circle is determined by the general
expression

\begin{equation}
Q=\int {{\bbox E}({\bbox r}){\bbox j}({\bbox r})\, d^2r},
\label{QQ}
\end{equation}
where ${\bbox E}({\bbox r})$ and ${\bbox j}({\bbox r})$ are the local
electric field and the local current density, respectively, at point
${\bbox r}$ inside of the circle. Since $\nabla\cdot{\bbox j}=0,
\nabla\times{\bbox E}=0$, we can express ${\bbox j}$ in terms of the
pseudoscalar $\psi$, and ${\bbox E}$ in terms of the electric
potential $\phi$, as given by
\begin{equation}
 {\bbox j}({\bbox r})=[\hat{\bbox z}\times\nabla\psi({\bbox r})], \ \
{\bbox E}({\bbox r})=-\nabla\phi({\bbox r}).
\label{psi1}
\end{equation}
In this notation, the three currents $I, I_1, I_2$ can be written as

\begin{equation}
I=\psi_1-\psi_3,\ \ I_1=\psi_2-\psi_1,\ \ I_2=\psi_2-\psi_3.
\label{3I}
\end{equation}
Substituting Eqs. (\ref{psi1}) into (\ref{QQ}), changing order in the mixed
product, and integrating by parts  we get

\begin{equation}
Q=-\int{\psi (\nabla\phi\cdot d\bbox{l})},
\label{QQ2}
\end{equation}
where the integral is taken along the closed circle in Fig. 2b. The circle
can be broken into three segments: 3-2, 2-1, and 1-3. Within each of the
segments we have  $\nabla\phi=(1/\sigma_i)\nabla\psi$, where the Hall
conductivity $\sigma_i$ is equal to $\sigma_2, \sigma_1$, and $\infty$ (metal),
respectively. Integration within separate segments yields

\begin{equation}
Q=-\frac{1}{2\sigma_2}(\psi_2^2-\psi_3^2)-
\frac{1}{2\sigma_1}(\psi_1^2-\psi_2^2).
\label{QQ3}
\end{equation}
Using Eqs. (\ref{I1I2}) and (\ref{3I}), we obtain finally

\begin{equation}
Q=\frac{I^2}{2(\sigma_2-\sigma_1)}\equiv R_3I^2,
\label{QQfin}
\end{equation}
which yields the formula (\ref{R3}) quoted in the main text.

Note, that if the colors in Fig. 2b are interchanged, which
corresponds to interchanging $\sigma_2$ and $\sigma_1$ in
Eq. (\ref{QQfin}), the Joule heat will be negative (recall that we
assume $\sigma_2 >\sigma_1$ everywhere in the paper). Since this would
contradict the Second Law of thermodynamics, for such a vertex, all
currents must be zero. Hence, only the vertices for which the
rightmost color, as shown in Fig. 2b, is white, can be ``active''
(can participate in the current transfer).

To illustrate our result (\ref{R3}), consider the simplest 
(without branching) example
 of a 3-vertex presented by a corner of a
rectangular homogeneous conductor with the Hall conductivity $\sigma_2$.  
Two metallic probes are attached at the bottom and at the top of the
rectangle, the sample on the left and on the right bordering to vacuum, $\sigma_1=0$.
If $\sigma_2 >0$,  the current will leave and enter
metallic probes focusing at the lower left and the upper right corners of the
sample which represent the active 3-vertices.  The two-terminal resistance
of such a
sample is equal, as easy to see, to the inverse Hall conductivity $1/\sigma_2$.  
This amounts to the effective resistance $R_3=1/2\sigma_2$
per each active 3-vertex, in agreement with Eq. (\ref{R3}).

As is easy to check, the same rule of selection of active vertices
applies if the metal in Fig. 2b is replaced by a vacuum, as is
appropriate for a Hall bar with an abrupt edge (Sec. IV). In contrast
to the edge to metal discussed above, we now have zero current out of
the edge, $I=0$, and a non-zero voltage difference $V=\phi_3-\phi_1$.
The expression for the Joule heat $Q$ can be obtained in almost the
same manner as Eq.  (\ref{QQfin}), except it is now convenient to use,
instead of Eq. (\ref{QQ2}), an equivalent formula

\begin{equation}
Q=\int{\phi (\nabla\psi\cdot d{\bbox l})}.
\label{QQQ}
\end{equation}
The final answer has the form

\begin{equation}
Q=\frac{V^2}{2(\sigma_1^{-1}-\sigma_2^{-1})},
\end{equation}
which has the same sign as the right-hand side of Eq. (\ref{QQfin}).

\begin{figure}
\caption{(a)~Local conductivity tensor components $\sigma_{xx}$ and
$\sigma_{xy}$ versus the local filling factor $\nu$. (b)~Macroscopic
fluctuations of the local filling factor. (c)~Conductivity tensor
distribution in inhomogeneous system at low temperatures. Black and
white regions correspond to quantized Hall regions with
$\sigma_{xy}=\sigma_1$ and $\sigma_2$, respectively, and
$\sigma_{xx}\approx 0$. The grey color which corresponds to the grey
strip in (b) shows the intermediate (non-quantized) region.}
\end{figure}

\begin{figure}
\caption{(a)~Conductivity distribution in an inhomogeneous Corbino
disc. Black and white are quantized regions as in Fig. 1(c), the
intermediate grey region is not shown. (b)~Complex ``fork vertex''
from the small square in (a) shown in magnified view. The arrows
schematically show currents passing between different regions by
focusing at simple vertices.}
\end{figure}

\begin{figure}
\caption{The temperature dependence of peak values of $\sigma_{xx}$
for different transitions between adjacent IQHE states obtained by
Rokhinson {\it et al.}\protect\cite{RokhimsonGoldman}. Different
symbols correspond to transitions: 5--6 ($\bigtriangledown$), 6--7 
($\diamond$), 7--8 ($\bullet$), 10--11 ($\times$), 13--14 ($\triangle$).
Some transitions are omitted by authors of Ref.~\protect\onlinecite{RokhimsonGoldman}
for sake of clarity. The geometric aspect ratios of the samples are
$A=0.21$ (sample A) and $A=0.32$ (sample B).}
\end{figure}

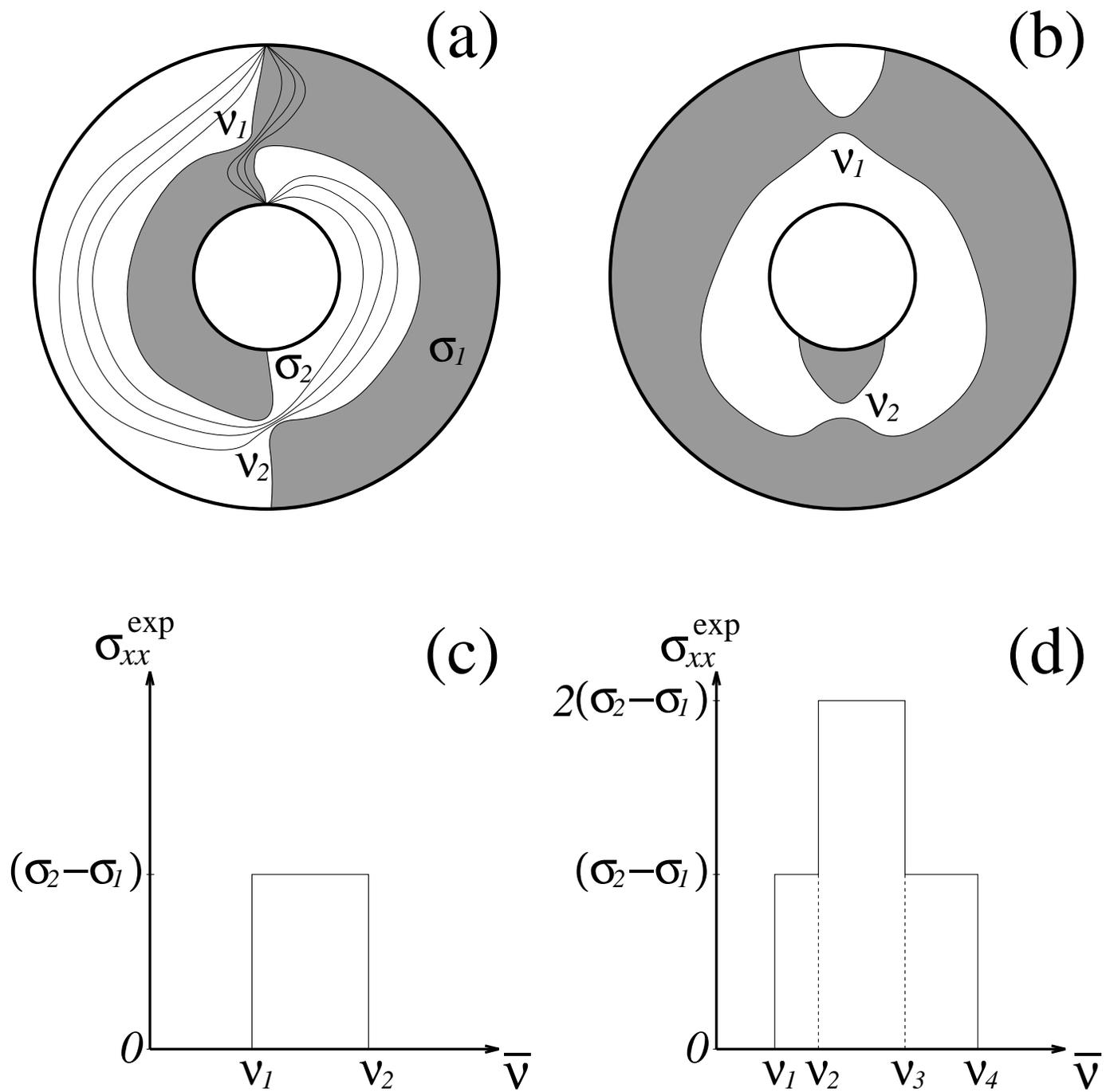
\begin{figure}
\caption{Two basic configurations of the conductivity distribution in
the Corbino disc in the saturation regime ($T\ll T_{s2}$) shown at
$\nu_1 < \overline\nu < \nu_2$, where $\nu_1$ and $\nu_2$ are filling
factors of the two saddle points. (a)~Configuration with a finite
two-terminal resistance. Thin lines are the lines of equal
potential. (b)~Configuration with infinite two-terminal resistance. No
current passing between contacts. (c)~Predicted observed diagonal conductance
$\sigma_{xx}^{\rm exp}$ {\it vs.} average filling factor
$\overline\nu$ for configuration (a).  (d)~Analogous dependence for a
configuration controlled by four saddle points.}
\end{figure}

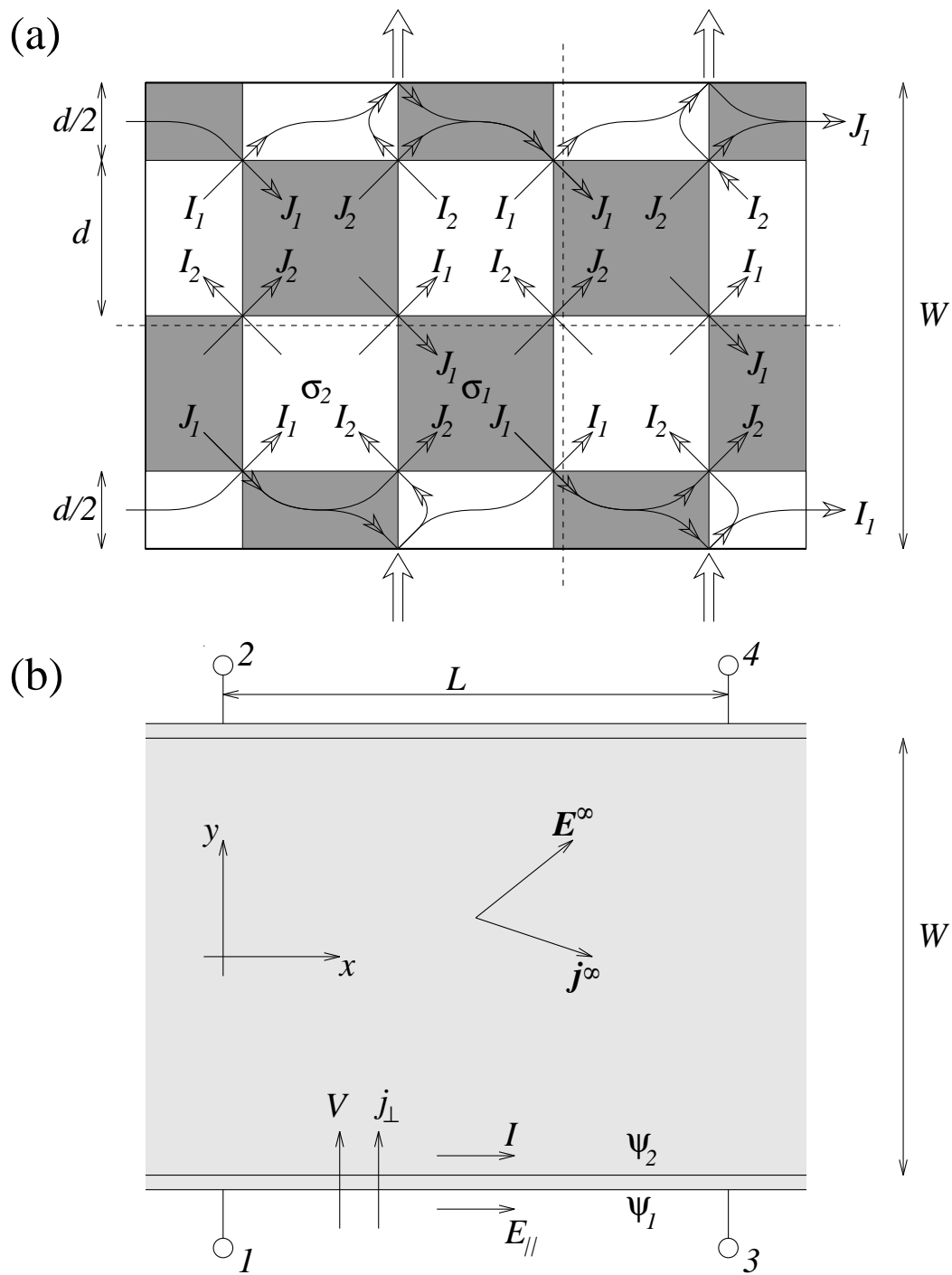
\begin{figure}
\caption{Diagram illustrating the boundary impedance matrix formalism.
(a)~Chessboard model of the two-phase distribution in a long
rectangular sample.  The arrows schematically show currents passing
from one (white or black) region to another by focusing at
vertices. Wide arrows show currents flowing from (to) metallic
contacts. (b)~Equivalent sample formed by infinitely thin boundary
strips attached to a homogeneous interior of the same width as that of
the sample in (a).}
\end{figure}


\end{document}